\documentclass[12pt]{article}

\catcode`@=11 \@addtoreset{equation}{section} \catcode `@=12

\def\be{\begin{equation}}
\def\ee{\end{equation}}

\begin{document}

\begin{center}
{\large\bf The quantum black hole in 2+1 dimensions} \\[.5cm]
{\large B. RAM$^{a,b}$, J. SHIRLEY$^b$} \\
$^a$Physics Department, New Mexico State University \\
Las Cruces, NM 88003, USA \\[.1cm]
$^b$Prabhu-Umrao Institute of Fundamental Research \\
A2/214 Janak Puri, New Delhi 110 058, India
\end{center}

\vspace*{2cm}

\begin{center}
Abstract
\end{center}
\bigskip

In this paper we investigate the quantum nature of a 2+1 dimensional
black hole using the method [arXiv: gr-qc/0504030] which earlier
revealed the quantum nature of a black hole in 3+1 dimensions.

\newpage

The classical analog in 2+1 dimensional Einstein theory with a
negative cosmological constant $\Lambda = {-1 \over \ell^2}$ of the
Schwarzschild line element was first obtained by Ba\~nados, Teitelboim
and Zanelli (BTZ) [1].  It is given by [2,3]
\be
ds^2 = \left(-8M + {r^2 \over \ell^2}\right) dt^2 - \left({dr^2 \over
\left(-8M + {r^2 \over \ell^2}\right)} + r^2 d\theta^2\right),
\label{one}
\ee
$M$ being the Arnowitt-Deser-Misner (ADM) mass.  Whereas the
Schwarzschild line element is asymptotically flat, (\ref{one}) is not.

The gravitational collapse of a disc of pressureless dust (the 2+1
dimensional analog of the Oppenheimer-Snyder collapse [4]) was studied
by Mann and Ross [5] who found that it -- the three spacetime $(3D)$
collapse -- exhibits properties that entirely parallel those of the
four spacetime $(4D)$ Oppenheimer-Snyder collapse.

The quantum nature of the $4D$ Schwarzschild black hole was revealed
in the paper by Ram, Ram and Ram [6] using a method in which one makes
quantum mechanics from time-like geodesics, obtaining thereby a
quantum equation which quantizes \underbar{mass}.  The use of Bose's
method [7] then reveals the quantum statistical nature of the $4D$
black hole, namely that it is a Bose-Einstein ensemble of quanta of
mass equal to twice the Planck mass.  In the present note we apply the 
method [6] to the 2+1 dimensional case.

The first integral of the time-like geodesics in three space-time
dimensions with the BTZ line element (\ref{one}) can be easily
obtained, and is given by
\be
\dot r^2 + L^2 \left(-{8M \over r^2} + {1 \over \ell^2}\right) - 8M +
{r^2 \over \ell^2} = E^2
\label{two}
\ee
with $E$ and $L$ constants of integration.  In Eq. (\ref{two}) dot
means differentiation with respect to the proper time.  Following
Ref. 6 we put $E = L = 0$ in (\ref{two}), and obtain 
\be
{1\over 2} \dot r^2 + {1\over 2} \omega^2 r^2 = 4M
\label{three}
\ee
with $\omega = {1\over\ell}$.  Proper use of the Schr\"odinger
prescription in (\ref{three}) then leads to the quantum equation
\be
\left(-{1\over 2r} {\partial \over \partial r} \left(r {\partial \over
\partial r}\right) + {1\over 2} \omega^2 r^2\right)\psi = 4 M \psi.
\label{four}
\ee
Thus the harmonic oscillator shows itself again, though in this case
it is a two-dimensional one as opposed to the four-dimensional one in
the $4D$ case.  With $U = r^{1/2} \psi$ and $M = \mu/4$,
Eq. (\ref{four}) takes the form
\be
\left[-{1\over2} \left({d^2 \over dr^2} + {1 \over 4r^2}\right) +
{1\over2} \omega^2 r^2\right]U = \mu U
\label{five}
\ee
with $\omega$ equal to two [8,9].  The quantum equation (\ref{five})
has the eigenvalues [10]
\be
\mu_n = 2(n+1)\omega, \ n = 0,1,2,\cdots .
\label{six}
\ee
Thus \underbar{mass} is quantized [11] as in (\ref{six}).  
That is, the nth
mass $(\mu)$ state is occupied by $n$ pairs of mass quanta, each
quantum being of $\omega = 2$ or of mass (energy) twice the Planck
mass (energy); same as in the $4D$ case.

With the establishment of the connection of the energy element of the
two-dimensional oscillator with the mass quantum of pure gravitation
via Eqs. (\ref{three}-\ref{six}), one can again use Bose's method to
calculate the entropy $S$ of pure gravitation within a $3D$ black
hole.  Let the mass of the BTZ black hole as in the paper of Mann and
Ross [5] be $M_B$ and let it contain $2N$ quanta.  Then $M_B =
2N\omega$.  Let us rewrite Eq. (\ref{six}) as
\be 
\epsilon_n = {\mu_n \over 2} = (n + 1) \omega, \ n = 0,1,2,\cdots,
\label{seven}
\ee
so that the nth $\epsilon$-state contains $n$ quanta.  The
thermodynamic quantity $E$ in Bose's paper [7] is given by
\be
E = {M_B \over 2} = N\omega,
\label{eight}
\ee
same as in the $4D$ case.  The entropy $S$ is again calculable using
the expression [6]
\be
S = {E \over T} - N\left(e^{\omega/T} - 1\right) \ell n\left(1 -
e^{-\omega/T}\right), 
\label{nine}
\ee
or
\be 
S = {M_B \over 2T} - {M_B \over 2\omega} \left[\left(e^{\omega/T} -
1\right) \ell n\left(1 - e^{-\omega/T}\right)\right].
\label{ten}
\ee
For ${\omega \over T} \ll 1$, i.e. for large $M_B$, Eq. (\ref{ten})
reduces to
\be
S = {M_B \over 2T} \left[1 - \ell n\left({\omega \over
T}\right)\right].
\label{eleven}
\ee

For the 2+1 dimensional black hole, if we take [2] $T = {r^+_B
\over 2\pi \ell^2} = {r^+_B \omega^2 \over 2\pi}$ with
$r^{+^2}_B = 8M_B \ell^2 = {8M_B \over \omega^2}$, then
Eq. (\ref{eleven}) becomes
\be
S = {2\pi r^+_B \over 4} \left[{1 - \ell n\left({\omega \over
T}\right) \over 4}\right] = {2\pi r^+_B \over 4} \left[{1 - \ell
n\left({\pi \over \sqrt{2M_B}}\right) \over 4}\right].
\label{twelve}
\ee
Eq. (\ref{twelve}) says that the entropy of a $3D$ black hole is given
by not just ${2\pi r^+_B \over 4}$ but by ${2\pi r^+_B \over4}$ times
a monotonically increasing function of the black hole mass $M_B$
[12,13].  This is the difference between the 2+1 dimensional black
hole and the 3+1 dimensional black hole for which the entropy comes
out to be ${A \over 4}$ ($A$ being the area of the $4D$ black hole and
$2\pi r^+_B$ being its $3D$ analog).

The above investigation demonstrates -- plainly -- that the method [6]
originally given for the $4D$ case works \underbar{as well} for the
3-dimensional case.  The investigation, moreover, brings out --
unambiguously -- the importance of quantum statistical effects.
\bigskip

\begin{center}
Acknowledgements
\end{center}
\smallskip

The authors thank N. Ram for a critical reading of the manuscript.
One of us (BR) thanks A. Ram, T. Halverson, R.S. Bhalerao, N. Banerjee
and R. Tipton for useful conversations, and TIFR and IUCAA for
pleasant stays.

\newpage

\begin{center}
References and Footnotes
\end{center}
\smallskip

\begin{enumerate}
\item[{[1]}] M. Ba\~nados, C. Teitelboim and J. Zanelli,
Phys. Rev. Lett. \underbar{69}, 1849 (1992).
\item[{[2]}] S. Carlip, Class. Quant. Grav. \underbar{22}, R85 (2005).
\item[{[3]}] Whereas in Ref. 1 the units used are such that $8G = 1 =
c$, we use units in which $G = 1 = c = \hbar = k$.
\item[{[4]}] J.R. Oppenheimer and H. Snyder, Phys. Rev. \underbar{56},
455 (1939).
\item[{[5]}] R.B. Mann and S.F. Ross, Phys. Rev. D\underbar{47}, 3319
(1993). 
\item[{[6]}] B. Ram, A. Ram, and N. Ram, arXiv: gr-qc/0504030; B. Ram,
Phys. Lett. A\underbar{265}, 1 (2000).
\item[{[7]}] S.N. Bose, Z. Phys. \underbar{26}, 178 (1924); English
translation in O. Theimer and B. Ram, Am. J. Phys. \underbar{44},
1056 (1976).
\item[{[8]}] Since Eq. (\ref{five}) with $\omega = 2$ is a map of [see
Ref. 6]
\[
\left[-{1\over2} \left({d^2 \over d\rho^2} + {1\over 4\rho^2}\right) -
{M \over \rho}\right] \phi(\rho) = - {1\over2} \phi(\rho).
\]
In this sense, $M$ is analog of the Schwarzschild mass; see also
Ref. 5. 
\item[{[9]}] V. Cardoso and J. Lemos take $\ell = 1$ in their paper
[Phys. Rev. D\underbar{63}, 124015 (2001)] which 
corresponds to $\omega = 1$.
\item[{[10]}] H.A. Mavromatis, \underbar{Exercises in Quantum
Mechanics} (D. Reidel Publishing Co., Dordrecht, Holland, 1986),
p. 95.
\item[{[11]}] In this connection, see also M.R. Setare,
Class. Quant. Grav. \underbar{21}, 1453 (2004).  We thank M. Setare
for bringing this paper to our attention.
\item[{[12]}] In this connection see p. 2869 of S. Carlip,
Class. Quant. Grav. \underbar{12}, 2853 (1995), and A. Ghosh and
P. Mitra, Mod. Phys. Lett. A\underbar{11}, 1231 (1996).
\item[{[13]}] For ${\omega \over T} = 10^{-20}$ or $M_B = 5 \times
10^{40}$, the numerical value of $\left[{1 - \ell n(\omega/T) \over
4}\right]$ is $\left[{1 + 20\ell n 10 \over 4}\right]$, and the
value increases monotonically by ${\ell n 10 \over 4}$ as ${\omega
\over T}$ decreases by a factor of 10 or $M_B$ increases by a factor
of 100.
\end{enumerate}

\end{document}